\documentclass[twocolumn,floatfix,superscriptaddress,showpacs,showkeys]{revtex4}
\usepackage{amsmath}
\usepackage{graphicx}
\usepackage{dcolumn}
\usepackage{bm}
\usepackage{booktabs}
\usepackage{amssymb}
\usepackage{multirow}
\usepackage{makecell}
\usepackage{textcomp}
\usepackage{subfigure}
\usepackage{phonetic}
\usepackage{extarrows}
\usepackage{color}
\usepackage{float}
\usepackage[colorlinks,citecolor=blue,linkcolor=blue]{hyperref}

\begin{document}


\title{Topological phase induced by distinguishing parameter regimes in cavity optomechanical system with multiple mechanical resonators}

\author{Lu Qi}
\affiliation{Department of Physics, Harbin Institute of Technology, Harbin, Heilongjiang 150001, China}
\author{Yan Xing}
\affiliation{Department of Physics, Harbin Institute of Technology, Harbin, Heilongjiang 150001, China}
\author{Shutian Liu}
\email{stliu@hit.edu.cn}
\affiliation{Department of Physics, Harbin Institute of Technology, Harbin, Heilongjiang 150001, China}
\author{Shou Zhang}
\email{szhang@ybu.edu.cn}
\affiliation{Department of Physics, Harbin Institute of Technology, Harbin, Heilongjiang 150001, China}
\affiliation{Department of Physics, College of Science, Yanbian University, Yanji, Jilin 133002, China}
\author{Hong-Fu Wang}
\email{hfwang@ybu.edu.cn}
\affiliation{Department of Physics, College of Science, Yanbian University, Yanji, Jilin 133002, China}


\date{\today}

\begin{abstract}
We propose two kinds of distinguishing parameter regimes to induce topological Su-Schrieffer-Heeger (SSH) phase in a one dimensional (1D) multi-resonator cavity optomechanical system via modulating the frequencies of both cavity fields and resonators. The introduction of the frequency modulations allows us to eliminate the Stokes heating process for the mapping of the tight-binding Hamiltonian without usual rotating wave approximation, which is totally different from the traditional mapping of the topological tight-binding model. We find that the tight-binding Hamiltonian can be mapped into a topological SSH phase via modifying the Bessel function originating from the frequency modulations of cavity fields and resonators, and the induced SSH phase is independent on the effective optomechanical coupling strength. On the other hand, the insensitivity of the system to the effective optomechanical coupling provides us another new path to induce the topological SSH phase based on the present 1D cavity optomechanical system. And we show that the system can exhibit a topological SSH phase via varying the effective optomechanical coupling strength in an alternative way, which is much more easier to be achieved in experiment. Furthermore, we also construct an analogous bosonic Kitaev model with the trivial topology by keeping the Stokes heating processes. Our scheme provides a steerable platform to investigate the effects of next-nearest-neighboring interactions on the topology of the system.   
\end{abstract}

\pacs{03.65.Vf, 73.43.Nq, 42.50.Wk, 07.10.Cm}
\keywords{topological phase, topological SSH model, cavity optomechanical system, frequency modulations}
\maketitle


\section{\label{sec.1}Introduction}
Over the past decades, cavity optomechanical system~\cite{Kippenberg1172,favero2009optomechanics}, which is composed of mechanical and optical modes, is becoming a fast-developing and appealing field for the investigation of fundamental quantum physics on the macroscopic scale. In the context of optomechanical system, various questions have been explored, such as entanglement between mechanical modes and cavity fields~\cite{vitali2007optomechanical,genes2008emergence,nie2014generating}, normal-mode splitting~\cite{dobrindt2008parametric,groblacher2009observation}, optomechanical induced transparency~\cite{wu2015tunable,li2016parity}, squeezing of light or resonators~\cite{jahne2009cavity,kronwald2014dissipative,gu2013squeezing,gu2013generation,zhang2005cooling,wang2011ultraefficient}, cooling resonators via feedback control~\cite{wilson2015measurement}, etc. Specially, more and more attention has been focused on the modulated optomechanical systems~\cite{mari2009gently,li2011fast,farace2012enhancing,liao2016generation,liao2015enhancement,wang2016macroscopic,yin2017nonlinear} in recent years, in which abundant physical phenomena have been reported. A scheme of optomechanical cooling has been proposed by utilizing the periodical modulations of frequency and damping of resonator~\cite{bienert2015optomechanical}. Another  cooling scheme of breaking quantum backaction limit has also been proposed based on an optomechanical system by modulating the frequencies of both the optical and mechanical components~\cite{wang2018optomechanical}, in which the Stokes heating processes can be safely and completely suppressed. The method of frequency modulation provides an effective path to eliminate the Stokes heating terms perfectly, which is essential to the simulations of all kinds of topological matters.    

The multi-resonator cavity optomechanical system~\cite{ludwig2013quantum,heinrich2011collective}, as the assemblage of a set of single optomechanical system, is also widely used to investigate 
diverse quantum questions~\cite{chang2011slowing,xiong2015asymmetric,xuereb2012strong,safavi2011electromagnetically,tomadin2012reservoir,xuereb2014reconfigurable}. de~Moraes~Neto {\it et al.}~\cite{de2016quantum} designed a scheme for robust quantum state transfer based on a 1D optomechanical array by using the decoupling method. Akram {\it et al.}~\cite{akram2012photon} proposed a scheme to achieve the photon-phonon entanglement in coupled optomechanical arrays. Wan {\it et al.}~\cite{wan2017controllable} realized the controllable photon and phonon localization induced by path interference in an optomechanical Lieb lattice. The simulations of $Z_{2}$ topological insulator~\cite{qi2017simulating} and topological bosonic Majorana chains~\cite{xing2018controllable} were also illustrated based on a 1D optomechanical array. Especially, in these previous literatures, the derivation of the effective Hamiltonian in optomechanical chain mainly used the rotating wave approximation to remove the Stokes heating processes. The exploration of topological SSH phase in a 1D multi-resonators optomechanical system, in which the Stokes heating processes are eliminated via modulating the frequencies of cavity fields and resonators, is rarely investigated yet.  

In this paper, we propose a scheme to induce the topological SSH phase by dint of two different parameter regimes based on a 1D multi-resonator cavity optomechanical system with the time-dependent frequency modulations of both cavity fields and mechanical modes. We find that, after eliminating the Stokes heating terms via the Bessel function, the 1D tight-binding Hamiltonian can be obtained to induce a topological SSH phase. There exists two parameter regimes for inducing the SSH phase. One is the effective nearest-neighboring (NN) hopping strength, which is modulated to satisfy the staggered dimerized hopping strength of SSH model by means of Bessel function originating from frequency modulations. The other is the effective optomechanical coupling, which is varied alternately to induce a topological SSH phase after removing the resonant Stokes heating processes completely. Furthermore, we also propose to build a bosonic Kitaev model based on the present 1D multi-resonator cavity optomechanical system, in which the system possesses a continuous energy spectrum corresponding to arbitrary strength of analogous pairing terms. Besides, we find that the next-nearest-neighboring (NNN) interactions between two adjacent cavity fields can be flexibly adjusted and even completely suppressed, which provides a novel and controllable platform to investigate the effect of NNN hopping on the topology of the system.

The paper is organized as follows: In Sec.~\ref{sec.2}, we derive the effective linearized Hamiltonian of the 1D multi-resonator optomechanical system with the time-dependent frequency modulations. In Sec.~\ref{sec.3}, we remove the Stocks heating terms of the Hamiltonian and induce a topological SSH phase via two different kinds of parameter regimes. Subsequently, an analogous bosonic Kitaev models is obtained via frequency modulations. Also, we investigate the effect of NNN hopping on the topology of the system. Finally, a conclusion is given in Sec.~\ref{sec.4}.

\section{\label{sec.2}System and Hamiltonian}
\begin{figure}
	\centering
	\includegraphics[width=0.9\linewidth]{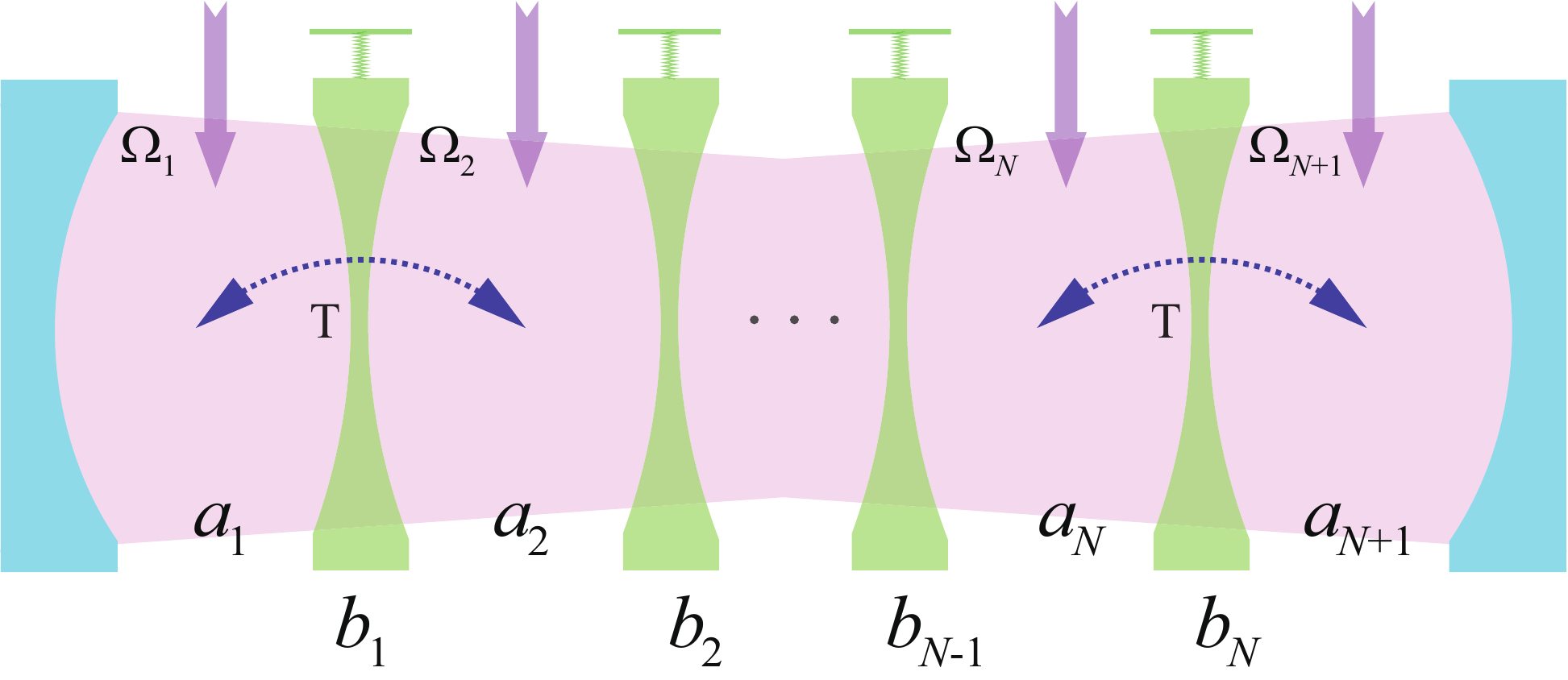}\\
	\caption{Schematic of 1D multi-resonator optomechanical system, which contains $N+1$ cavity modes and $N$ resonators. Each cavity mode is driven by a laser field and the coupling between resonator $b_{n}$ and cavity field $a_{n}$ ($a_{n+1}$) is $g_{n}$.}\label{fig1}
\end{figure}
Consider a 1D multi-resonator cavity optomechanical system composed of $N+1$ cavity fields and $N$ resonators (see Appendix for the case of $N$ cavity fields and $N$ resonators), in which the frequencies of cavity fields and resonators can be modulated, as shown in Fig.~\ref{fig1}. In this array, each cavity field is driven by a laser with frequency $\omega_{d}$ and strength $\Omega_{n}$. The two adjacent cavity fields possess direct coupling with hopping strength $T$ and the single-phonon optomechanical coupling strength between resonator $b_{n}$ and cavity field $a_{n}$ ($a_{n+1}$) is $g_{n}$. In this way the system is dominated by the following Hamiltonian
\begin{eqnarray}\label{e01}
H&=&\sum_{n=1}^{N+1}\left[\omega_{a,n}+\lambda_{n} \nu\cos(\nu t+\phi)\right]a_{n}^{\dag}a_{n}\cr\cr
&&+\sum_{n=1}^{N} \left[\omega_{b,n}+\gamma_{n} \nu\cos(\nu t+\phi)\right]b_{n}^{\dag}b_{n}\cr\cr
&&+\sum_{n=1}^{N+1}(\Omega_{n}a_{n}^{\dag}e^{-i\omega_{d}t}+\Omega_{n}^{\ast}a_{n}e^{i\omega_{d}t})\cr\cr
&&-\sum_{n=1}^{N}g_{n} (a_{n}^{\dag}a_{n}-a_{n+1}^{\dag}a_{n+1})(b_{n}^{\dag}+b_{n})\cr\cr
&&+\sum_{n=1}^{N}T(a_{n+1}^{\dag}a_{n}+a_{n}^{\dag}a_{n+1}),
\end{eqnarray}
where $a_{n}^{\dag}$ $(b_{n}^{\dag})$ is the creation operator of optical cavity field (mechanical resonator) and $a_{n}$ $(b_{n})$ is its corresponding annihilation operator. The first two terms represent the modulated free energy of cavity fields and resonators with the modulated strength $\lambda_{n}$ ($\gamma_{n}$), frequency $\nu$, and phase $\phi$. The frequency modulations of cavity field and resonator can be realized experimentally via a laser irradiated into the cavity field~\cite{1994APL2877} and a gate electrode applied on the resonator~\cite{Singh2014}, respectively. The third term describes the interaction between cavity field and external driving laser field. The fourth term represents the interaction between cavity field $a_{n}$ ($a_{n+1}$) and mechanical resonator $b_{n}$. And the last term denotes the direct interaction between two adjacent cavity fields.

Under the condition of strong laser driving, we choose to work in the rotating frame with respect to the driving frequency $\omega_{d}$ and rewrite the operators as $a_{n}=\alpha_{n}+\delta a_{n}$ ($b_{n}=\beta_{n}+\delta b_{n}$). After dropping the notation $\delta$ for all the fluctuation operators $\delta a_{n}$ ($\delta b_{n}$), the Hamiltonian is given by
\begin{eqnarray}\label{e02}
H_{L}&=&\sum_{n=1}^{N+1}[\Delta_{a,n}^{'}+\lambda_{n} \nu\cos(\nu t+\phi)]a_{n}^{\dag}a_{n}\cr\cr
&&+\sum_{n=1}^{N} [\omega_{b,n}+\gamma_{n} \nu\cos(\nu t+\phi)]b_{n}^{\dag}b_{n}\cr\cr
&&-\sum_{n=1}^{N}g_{n} (\alpha_{n}^{\ast}a_{n}+\alpha_{n}a_{n}^{\dag}\cr\cr
&&-\alpha_{n+1}^{\ast}a_{n+1}-\alpha_{n+1}a_{n+1}^{\dag})(b_{n}^{\dag}+b_{n})\cr\cr
&&+\sum_{n=1}^{N}T(a_{n+1}^{\dag}a_{n}+a_{n}^{\dag}a_{n+1}),
\end{eqnarray}
where $\Delta_{a,n}^{'}$ is effective detuning caused by optomechanical coupling with $\Delta_{a,1}^{'}=\Delta_{a,1}-g_{1}[\beta_{1}^{\ast}+\beta_{1}]$, $\Delta_{a,N+1}^{'}=\Delta_{a,N+1}+g_{N}[\beta_{N}^{\ast}+\beta_{N}]$, $\Delta_{a,n=2...N}^{'}=\Delta_{a,n}-g_{n-1}(\beta_{n-1}^{\ast}+\beta_{n-1})+g_{n}(\beta_{n}^{\ast}+\beta_{n})$, and $\Delta_{a,n}=\omega_{a,n}-\omega_{d}$ is the detunings between cavity fields and driving fields. After performing the rotating transformation defined by 
\begin{eqnarray}\label{e03}
V&=&\mathrm{exp}\bigg\{\sum_{n=1}^{N+1}-i\Delta_{a,n}^{'}ta_{n}^{\dag}a_{n}-i\lambda_{n}\sin(\nu t+\phi) a_{n}^{\dag}a_{n}\cr\cr
&&+\sum_{n=1}^{N}-i\omega_{b,n}tb_{n}^{\dag}b_{n}-i\gamma_{n} \sin(\nu t+\phi) b_{n}^{\dag}b_{n}\bigg\},
\end{eqnarray}
the Hamiltonian becomes
\begin{eqnarray}\label{e04}
H_{L}^{'}&=&\sum_{n} \left\{-G_{n}a_{n}^{\dag}b_{n}e^{i[(\Delta_{a,n}^{'}-\omega_{b,n})t+(\lambda_{n}-\gamma_{n})\sin(\nu t+\phi) ]}\right.\cr\cr
&&\left.-G_{n}a_{n}^{\dag}b_{n}^{\dag}e^{i[(\Delta_{a,n}^{'}+\omega_{b,n})t+(\lambda_{n}+\gamma_{n})\sin(\nu t+\phi) ]}\right.\cr\cr
&&\left.+G_{n+1}a_{n+1}^{\dag}b_{n}e^{i[(\Delta_{a,n+1}^{'}-\omega_{b,n})t+(\lambda_{n+1}-\gamma_{n})\sin(\nu t+\phi) ]}\right.\cr\cr
&&\left.+G_{n+1}a_{n+1}^{\dag}b_{n}^{\dag}e^{i[(\Delta_{a,n+1}^{'}+\omega_{b,n})t+(\lambda_{n+1}+\gamma_{n})\sin(\nu t+\phi) ]}\right.\cr\cr
&&\left.+T a_{n+1}^{\dag}a_{n} e^{i(\lambda_{n+1}-\lambda_{n})\sin (\nu t+\phi)}\right\}+\mathrm{H.c.},
\end{eqnarray}
where $G_{n}=g_{n}\alpha_{n}$ ($G_{n+1}=g_{n}\alpha_{n+1}$) is the effective optomechanical coupling. Note that the first and the third terms in Eq.~(\ref{e04}) represent anti-Stokes terms of cooling the resonators while the second and the fourth terms in Eq.~(\ref{e04}) describe the Stokes heating processes of resonators. Apparently, the different forms of the frequency modulations of cavity fields and resonators have  significant influences on the system. 

\section{\label{sec.3}Topological phase and phase transition induced by different parameter regimes}
The crucial issue in the process of inducing an usual topological phase is to derive the tight-binding Hamiltonian, which implies that the Stokes heating terms should be eliminated. Here we use the frequency modulations to remove the Stokes heating terms and induce a topological SSH phase in terms of two different parameter regimes in the following section.

\subsection{\label{sec.A}Topological SSH phases induced by frequency modulations}
To obtain the necessary tight-binding Hamiltonian for the simulation of the standard SSH model, we consider the case that system does not possess the direct coupling between two adjacent cavity fields ($T=0$) and take the phase of frequency modulations as $\phi=0$. After exploiting the Jacobi$-$Anger expansions $e^{i\kappa \sin\nu t}=\sum_{m=-\infty}^{\infty}J_{m}(\kappa)e^{im\nu t}$, the Hamiltonian in Eq.~(\ref{e04}) can be rewritten as
\begin{widetext}
\begin{eqnarray}\label{e05}
H_{L,A_{1}}&=&\sum_{n} \left\{-\sum_{m_{1}=-{\infty}}^{\infty}G_{n}J_{m_{1}}(\kappa_{1,n})a_{n}^{\dag}b_{n}e^{i[(\Delta_{a,n}^{'}-\omega_{b,n})+m_{1}\nu]t }-\sum_{m_{2}=-{\infty}}^{\infty}G_{n}J_{m_{2}}(\kappa_{2,n})a_{n}^{\dag}b_{n}^{\dag}e^{i[(\Delta_{a,n}^{'}+\omega_{b,n})+m_{2}\nu]t }\right.\cr\cr
&&\left.+\sum_{m_{3}=-{\infty}}^{\infty}G_{n+1}J_{m_{3}}(\kappa_{3,n})a_{n+1}^{\dag}b_{n}e^{i[(\Delta_{a,n+1}^{'}-\omega_{b,n})+m_{3}\nu]t}+\sum_{m_{4}=-{\infty}}^{\infty}G_{n+1}J_{m_{4}}(\kappa_{4,n})a_{n+1}^{\dag}b_{n}^{\dag}e^{i[(\Delta_{a,n+1}^{'}+\omega_{b,n})+m_{4}\nu]t}\right\}\cr\cr
&&+\mathrm{H.c.},
\end{eqnarray}
\end{widetext}
where $\kappa_{1,n}=\lambda_{n}-\gamma_{n}$, $\kappa_{2,n}=\lambda_{n}+\gamma_{n}$, $\kappa_{3,n}=\lambda_{n+1}-\gamma_{n}$,  $\kappa_{4,n}=\lambda_{n+1}+\gamma_{n}$, and $J_{m_{j}}(\kappa_{j,n})$ is the $m_{j}$th order of the first kind of Bessel function with $j=1, 2, 3, 4 $.

When the frequency modulation parameters satisfy $\Delta_{a,n}^{'}=\Delta_{a,n+1}^{'}=\omega_{b,n}$, $\nu=\omega_{b,n}$, $m_{1}=m_{3}=0$, and $m_{2}=m_{4}=-2$, we find that the system possesses resonant anti-Stokes terms and Stokes heating terms simultaneously. Then the Hamiltonian in Eq.~(\ref{e05}) becomes 
\begin{eqnarray}\label{e06}
H_{L,A_{2}}&&=\sum_{n} \left\{-G_{n}J_{0}(\kappa_{1,n})a_{n}^{\dag}b_{n}-G_{n}J_{-2}(\kappa_{2,n})a_{n}^{\dag}b_{n}^{\dag}\right.\cr\cr
&&\left.+G_{n+1}J_{0}(\kappa_{3,n})a_{n+1}^{\dag}b_{n}+G_{n+1}J_{-2}(\kappa_{4,n})a_{n+1}^{\dag}b_{n}^{\dag}\right\}\cr\cr
&&+\mathrm{H.c.}.
\end{eqnarray}
\begin{figure}
	\centering
	\subfigure{\includegraphics[width=0.24\linewidth]{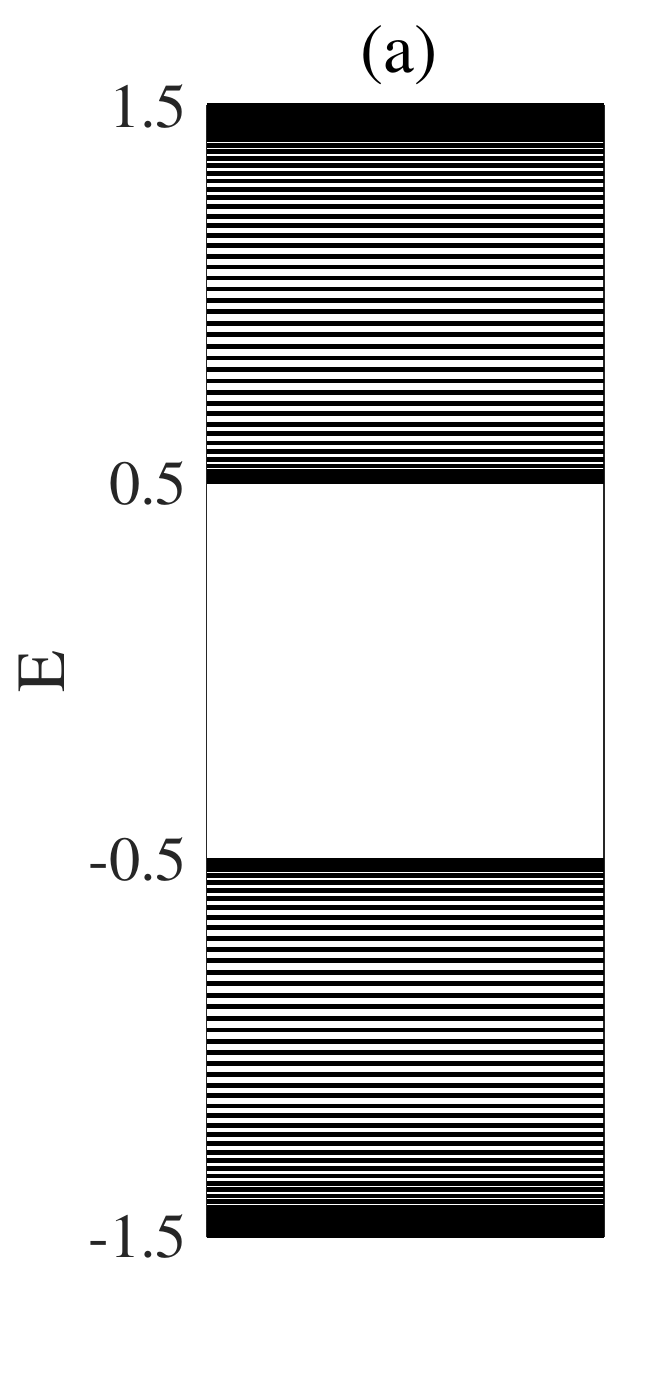}}
	\hspace{-2mm}
	\subfigure{\includegraphics[width=0.24\linewidth]{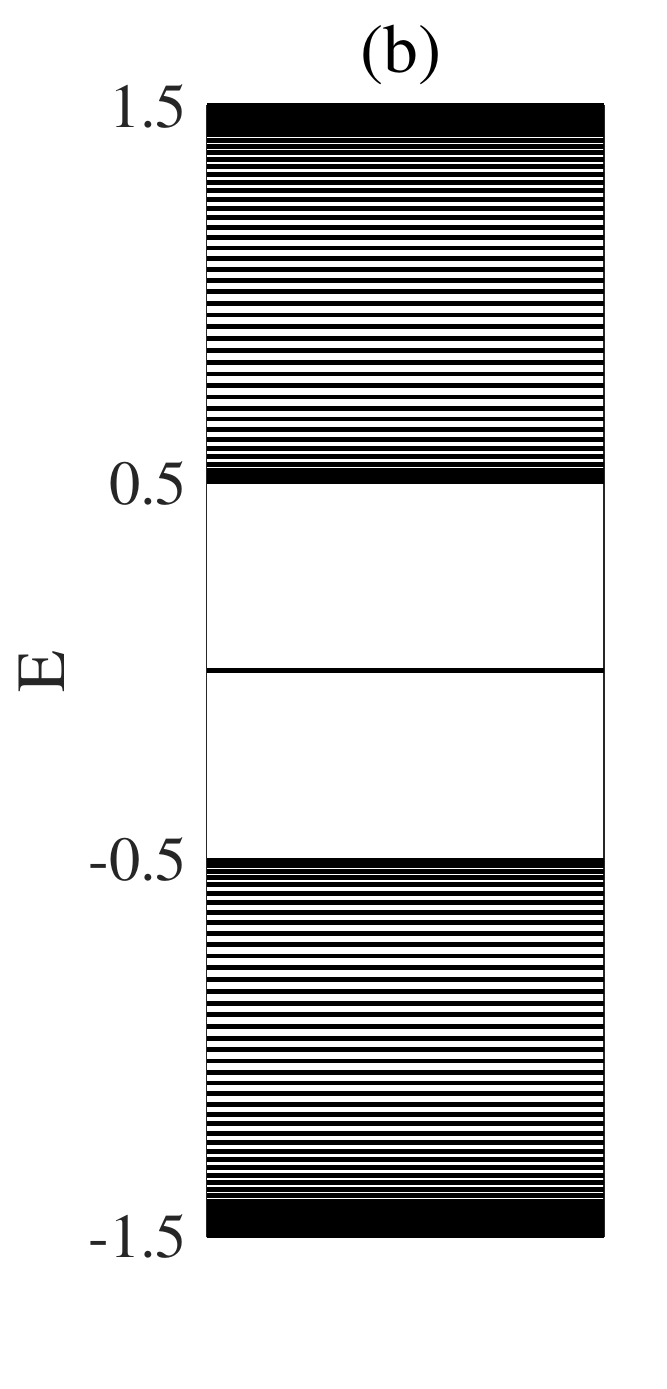}}
	\hspace{-2mm}
	\subfigure{\includegraphics[width=0.24\linewidth]{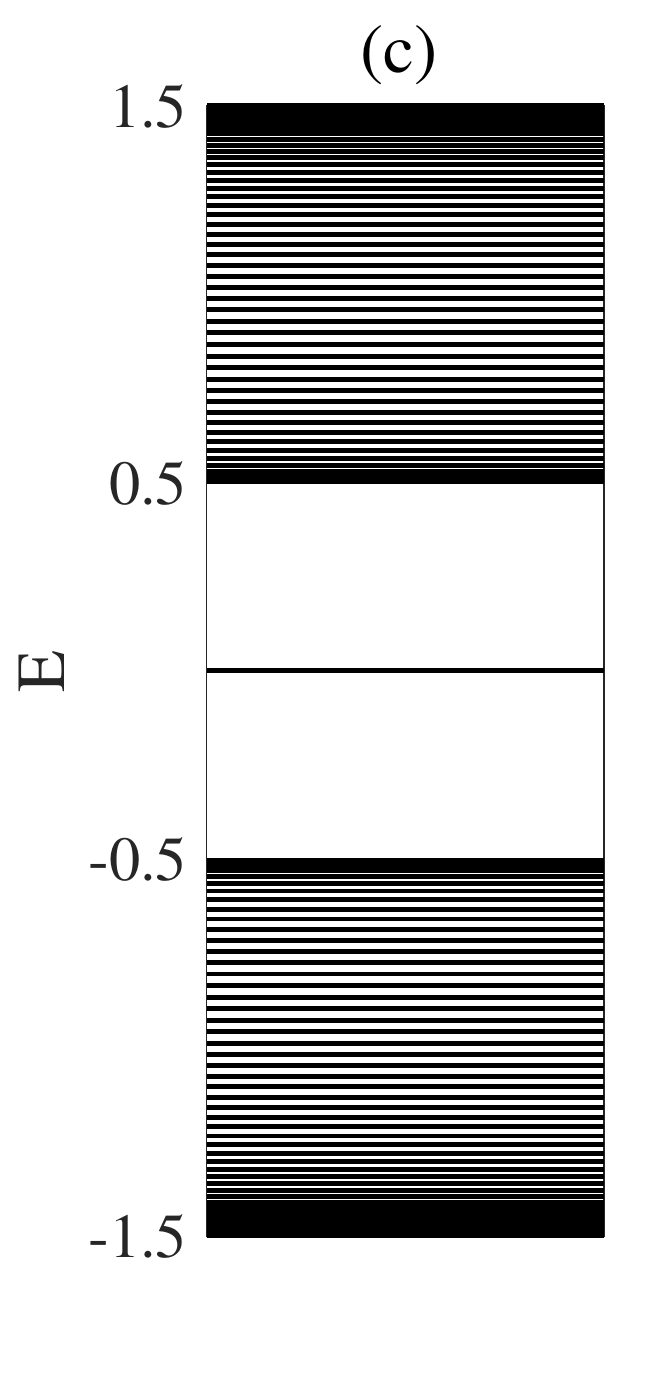}}
	\hspace{-2mm}
	\subfigure{\includegraphics[width=0.24\linewidth]{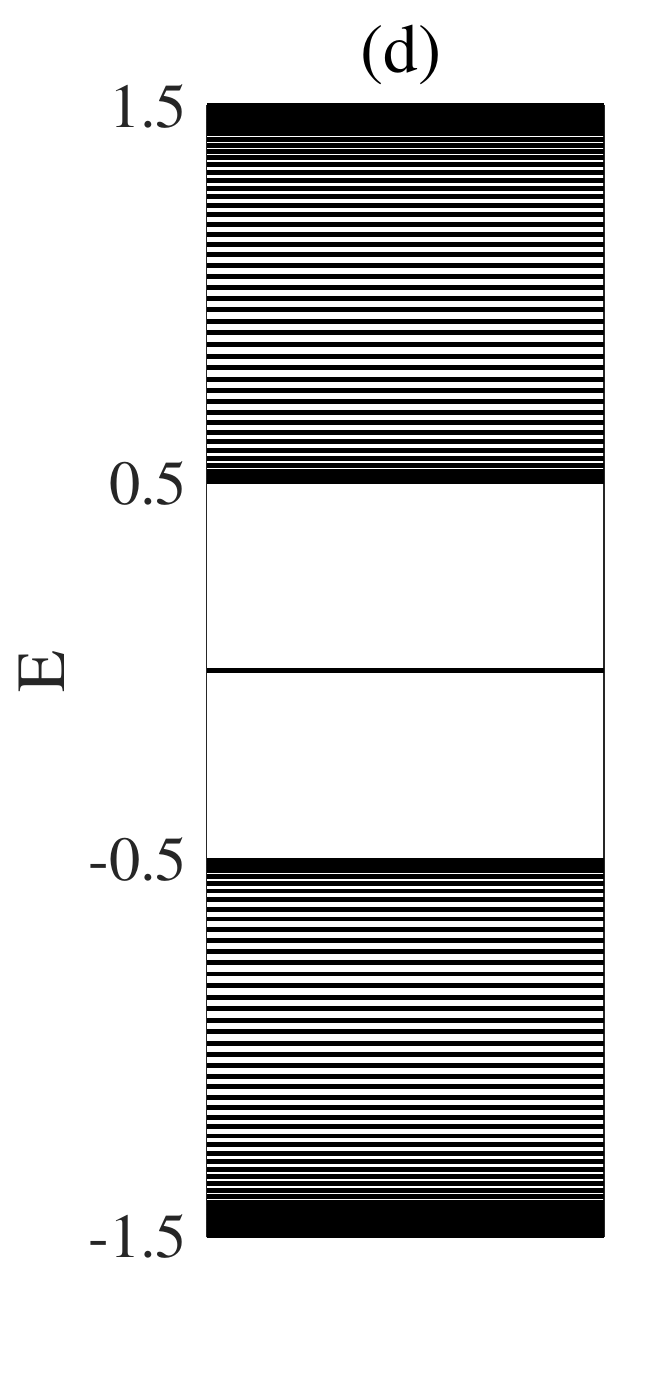}} 
	\subfigure{\includegraphics[width=1.0\linewidth]{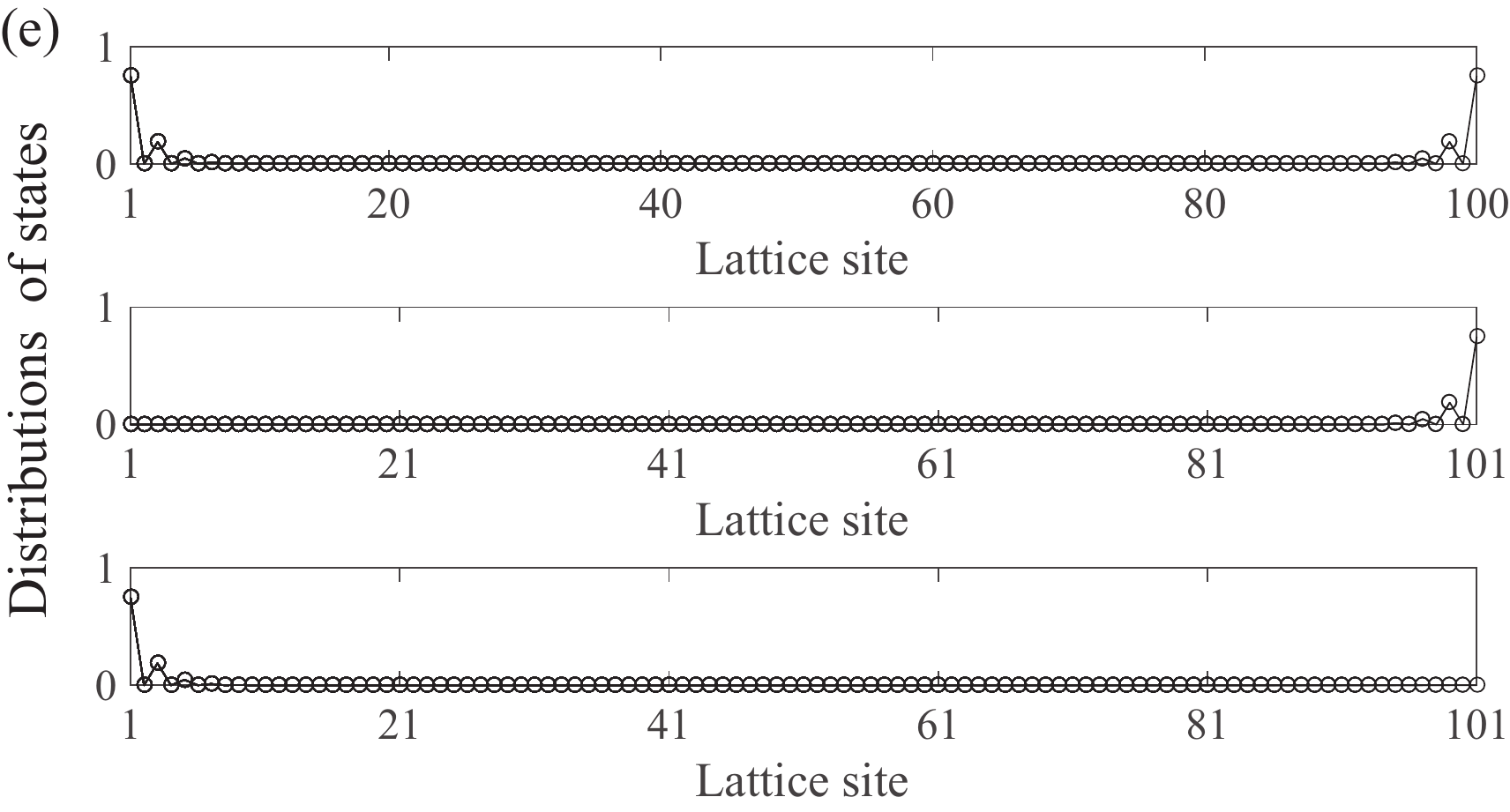}}
	
	\subfigure{\includegraphics[width=0.9\linewidth]{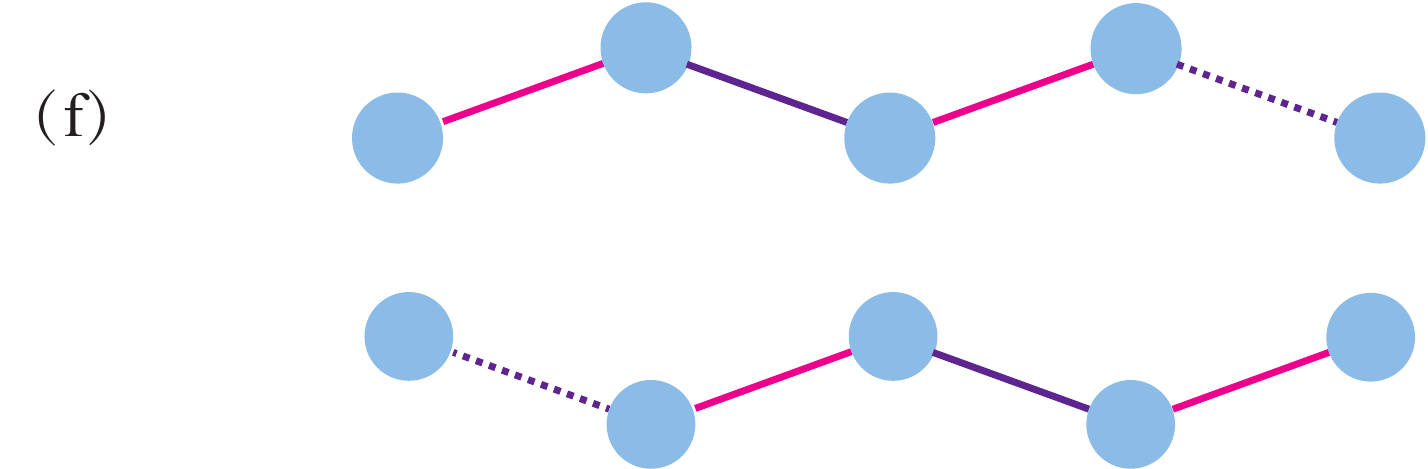}} 
	\caption{The energy spectra and distributions of edge states. (a) Energy spectrum corresponding to $J_{0}(\kappa_{1,n})>J_{0}(\kappa_{3,n})$, and the size of cavity optomechanical lattice chain system is $L=100$. (b) Energy spectrum corresponding to $J_{0}(\kappa_{1,n})<J_{0}(\kappa_{3,n})$, $L=100$. (c) Energy spectrum corresponding to $J_{0}(\kappa_{1,n})>J_{0}(\kappa_{3,n})$, $L=101$. (d) Energy spectrum corresponding to $J_{0}(\kappa_{1,n})<J_{0}(\kappa_{3,n})$, $L=101$. (e) The distributions of edge states in (b), (c), and (d). (f) The schematic diagram of odd-even effect of SSH chain. The red lines, purple lines, and dashed lines represent the strong hopping bonds, weak hopping bonds, and the decoupling of edge site, respectively. When the size of cavity optomechanical lattice chain system is an odd number, the last site will decouple from the lattice chain when $-G_{n}J_{-1}(\kappa_{1,n})>G_{n+1}J_{-1}(\kappa_{3,n})$ and the first site will separate from the lattice chain when $-G_{n}J_{-1}(\kappa_{1,n})<G_{n+1}J_{-1}(\kappa_{3,n})$. The other parameters are taken as $-G_{n}=G_{n+1}$.}\label{fig2}
\end{figure}

Obviously, the resonant Stokes heating terms in Eq.~(\ref{e06}) are useless in the usual mapping of tight-binding Hamiltonian. In the previous investigations with respect to the mappings of bosonic topological tight-binding Hamiltonian, the Stokes heating terms are removed mainly by using the rotating wave approximation~\cite{qi2017simulating,mei2015simulation,mei2016witnessing}. Here we apply frequency modulation method to dealing with the Stokes heating processes. By choosing the Bessel coefficients to satisfy $J_{-2}(\kappa_{2,n})=J_{-3}(\kappa_{4,n})=0$, the Stokes heating terms
can be suppressed and even eliminated completely, and the tight-binding Hamiltonian is expressed as 
\begin{eqnarray}\label{e07}
H_{L,A}&=&\sum_{n} \left[G_{n}J_{0}(\kappa_{1,n})a_{n}^{\dag}b_{n}+G_{n+1}J_{0}(\kappa_{3,n})a_{n+1}^{\dag}b_{n}\right]\cr\cr
&&+\mathrm{H.c.}.
\end{eqnarray} 
Obviously, the above Hamiltonian is equivalent to a SSH-type Hamiltonian if we map the cavity field $a_{n}$ and resonator $b_{n}$ as two sites of the SSH model.
Theoretically, the condition of $J_{-2}(\kappa_{2,n})=J_{-2}(\kappa_{4,n})=0$ can be easily achieved via choosing the parameters to satisfy $\kappa_{2,n}\sim\infty$, $\kappa_{4,n}\sim\infty$, and $\kappa_{2,n}\approx \kappa_{4,n}$. In experiment, we can choose the strong enough strength of frequency modulations to make the Bessel coefficients satisfy $J_{-2}(\kappa_{2,n})=J_{-2}(\kappa_{4,n})=0$ approximately. The previous investigations about inducing the topological phase based on the 1D multi-resonator optomechanical system mainly depend on the periodic modulation of effective optomechanical coupling strength~\cite{qi2017simulating}. We here propose a new way to induce a topological SSH phase via Bessel function originating from the frequency modulations of cavity fields and resonators. The condition of strong enough strength of frequency modulations ($\kappa_{2,n}\approx\kappa_{4,n}\sim\infty$)  means that the other two Bessel parameters should satisfy $\kappa_{1,n}\ne \kappa_{3,n}$. Thus we can obtain different values of the zeroth Bessel function (such as $J_{0}(\kappa_{1,n})<J_{0}(\kappa_{3,n})$ or $J_{0}(\kappa_{1,n})>J_{0}(\kappa_{3,n})$) via choosing the different values of $\kappa_{1,n}$ and $\kappa_{3,n}$. To emphasize the impact of Bessel coefficients $J_{0}(\kappa_{1,n})$ and $J_{0}(\kappa_{3,n})$, we take the effective optomechanical coupling strength to satisfy $-G_{n}=G_{n+1}$ (for simplicity, we choose the values of $-G_{n}$ and $G_{n+1}$ to be positive and real in all of the following contents). Then the topological properties of the system can be identified by the values of $J_{0}(\kappa_{1,n})$ and $J_{0}(\kappa_{3,n})$. When the Bessel function is chosen to satisfy $J_{0}(\kappa_{1,n})>J_{0}(\kappa_{3,n})$, the multi-resonator optomechanical system exhibits a topologically trivial phase corresponding to a lattice chain with even number sites, as revealed in Fig.~\ref{fig2}(a). Conversely, when it satisfies $J_{0}(\kappa_{1,n})<J_{0}(\kappa_{3,n})$, the system possesses two degenerate edge states located at both ends of the system with even number sites, as shown in Fig.~\ref{fig2}(b). For the case that the size of the multi-resonator optomechanical system is taken as an odd number, we find that the system possesses an edge state located at the rightmost end ($J_{0}(\kappa_{1,n})>J_{0}(\kappa_{3,n})$) or the leftmost end ($J_{0}(\kappa_{1,n})<J_{0}(\kappa_{3,n})$) of the optomechanical array, which is induced by the odd-even effect of SSH model, as shown in Figs.~\ref{fig2}(c) and~\ref{fig2}(d). To further clarify the topological edge states, we plot the distributions of edge state corresponding to an even number and an odd number of lattice size, respectively, as shown in Fig.~\ref{fig2}(e). Also, the schematic diagram of odd-even effect of SSH chain is depicted in Fig.~\ref{fig2}(f). 

We should stress that the topological SSH phase can also be induced corresponding to $-G_{n}\ne G_{n+1}$ and even with $-G_{n}>G_{n+1}$. Due to the selectivity of the values of Bessel function, we can always find a set of $J_{0}(\kappa_{1,n})$ and $J_{0}(\kappa_{3,n})$ to make the effective hopping parameters between adjacent cavity field and resonator satisfy $-G_{n}J_{0}(\kappa_{1,n})<G_{n+1}J_{0}(\kappa_{3,n})$ in the certain range even when $-G_{n}>G_{n+1}$. This weak-strong alternating effective hopping strength is the crucial condition to induce the SSH phase. Apparently, the method we use to induce the SSH phase is independent on the effective optomechancal coupling strength. The insensitivity of the present scheme on the optomechancal coupling strength provides a new path to induce topological SSH phase in 1D multi-resonator optomechanical system. Meanwhile, we clarify that this method needs the strong enough strength of frequency modulations, which may exist certain difficulties in experiment. In order to avoid this obstacle on the experimental simulation of the nontrivial SSH phase based on the present cavity optomechanical system, we propose another parameter regime to induce SSH phase in the next subsection.

\subsection{\label{sec.B}Topological SSH phases induced via strong effective optomechanical coupling}
Here we propose how to eliminate the resonant Stokes heating terms via the method of effective optomechanical coupling parameter regime and show that the topological phase can be induced successfully and effectively. Choosing the modulation strengths of cavity field and resonator to satisfy $\lambda_{n+1}=\lambda_{n}=\gamma_{n}$ and $T=\phi=0$, then the Hamiltonian in Eq.~(\ref{e04}) becomes
\begin{eqnarray}\label{e08}
H_{L,B_{1}}&=&\sum_{n} \left\{-G_{n}a_{n}^{\dag}b_{n}e^{i(\Delta_{a,n}^{'}-\omega_{b,n})t}\right.\cr\cr
&&\left.-G_{n}a_{n}^{\dag}b_{n}^{\dag}e^{i[(\Delta_{a,n}^{'}+\omega_{b,n})t+2\lambda_{n}\sin(\nu t) ]}\right.\cr\cr
&&\left.+G_{n+1}a_{n+1}^{\dag}b_{n}e^{i(\Delta_{a,n+1}^{'}-\omega_{b,n})t}\right.\cr\cr
&&\left.+G_{n+1}a_{n+1}^{\dag}b_{n}^{\dag}e^{i[(\Delta_{a,n+1}^{'}+\omega_{b,n})t+(\lambda_{n+1}+\lambda_{n})\sin(\nu t) ]}\right\}\cr\cr
&&+\mathrm{H.c.}.
\end{eqnarray} 
After exploiting the Jacobi$-$Anger expansions, the Hamiltonian is expressed as
\begin{widetext}
\begin{eqnarray}\label{e09}
H_{L,B_{2}}&=&\sum_{n} \left\{-G_{n}a_{n}^{\dag}b_{n}e^{i(\Delta_{a,n}^{'}-\omega_{b,n})t}+G_{n+1}a_{n+1}^{\dag}b_{n}e^{i(\Delta_{a,n+1}^{'}-\omega_{b,n})t}+\sum_{m_{1}=-\infty}^{\infty} -G_{n}J_{m_{1}}(\kappa_{1,n})a_{n}^{\dag}b_{n}^{\dag}e^{i[(\Delta_{a,n}^{'}+\omega_{b,n})+m_{1}\nu ]t}\right.\cr\cr
&&\left.+\sum_{m_{2}=-\infty}^{\infty}G_{n+1}J_{m_{2}}(\kappa_{2,n})a_{n+1}^{\dag}b_{n}^{\dag}e^{i[(\Delta_{a,n+1}^{'}+\omega_{b,n})+m_{2}\nu ]t}\right\}+\mathrm{H.c.},
\end{eqnarray}
\end{widetext}
where $\kappa_{1,n}=2\lambda_{n}$ and $\kappa_{2,n}=\lambda_{n+1}+\lambda_{n}$. 
Under the red-detuning regime of $\Delta_{a,n}^{'}=\Delta_{a,n+1}^{'}=\omega_{b,n}$,
the time-dependent exponential in NN hopping terms can be removed safely, the Hamiltonian can be rewritten as 
\begin{eqnarray}\label{e10}
H_{L,B_{3}}&=&\sum_{n} \left[-G_{n}a_{n}^{\dag}b_{n}+G_{n+1}a_{n+1}^{\dag}b_{n}\right.\cr\cr
&&\left.+\sum_{m_{1}=-\infty}^{\infty} -G_{n}J_{m_{1}}(\kappa_{1,n})a_{n}^{\dag}b_{n}^{\dag}e^{i(2\omega_{b,n}+m_{1}\nu )t}\right.\cr\cr
&&\left.+\sum_{m_{2}=-\infty}^{\infty}G_{n+1}J_{m_{2}}(\kappa_{2,n})a_{n+1}^{\dag}b_{n}^{\dag}e^{i(2\omega_{b,n}+m_{2}\nu )t}\right]\cr\cr
&&+\mathrm{H.c.}.
\end{eqnarray} 
Thus the system has resonant Stokes heating terms when $\nu=\omega_{b,n}$ and $m_{1}=m_{2}=-2$, leading that 
\begin{eqnarray}\label{e11}
H_{L,B_{4}}&=&\sum_{n} \left[-G_{n}a_{n}^{\dag}b_{n}+G_{n+1}a_{n+1}^{\dag}b_{n}\right.\cr\cr
&&\left.-G_{n}J_{-2}(\kappa_{1,n})a_{n}^{\dag}b_{n}^{\dag}\right.\cr\cr
&&\left.+G_{n+1}J_{-2}(\kappa_{2,n})a_{n+1}^{\dag}b_{n}^{\dag}\right]+\mathrm{H.c.}.
\end{eqnarray}  
Apparently, we can safely remove the resonant Stokes heating terms by choosing appropriate values of $\kappa_{1,n}$ and $\kappa_{2,n}$ to satisfy $J_{-2}(\kappa_{1,n})=J_{-2}(\kappa_{2,n})=0$. We stress that the process of eliminating Stokes heating terms does not involve any rotating wave approximation. Thus this method provides a new way to obtain the tight-binding Hamiltonian for inducing topological phase based on the present optomechanical system with frequency modulations. Another advantage is that the effective optomechanical coupling $G_{n}$ ($G_{n+1}$) can realize the strong coupling regime, in which the couplings between resonator and the adjacent cavity fields can be selected in a wider range. Then the system can be described by the following tight$-$binding SSH Hamiltonian
\begin{eqnarray}\label{e12}
H_{L,B}&=&\sum_{n} \bigg[-G_{n}a_{n}^{\dag}b_{n}+G_{n+1}a_{n+1}^{\dag}b_{n}\bigg]+\mathrm{H.c.}.  
\end{eqnarray}  
Based on this Hamiltonian, the topologically nontrivial SSH phase can be obtained by varying the adjacent effective optomechanical coupling strength in an alternative way, such as $-G_{n}<G_{n+1}$.

\subsection{\label{sec.C}Topologically trivial Kitaev model}
\begin{figure}
	\centering
	\subfigure{\includegraphics[width=0.85\linewidth]{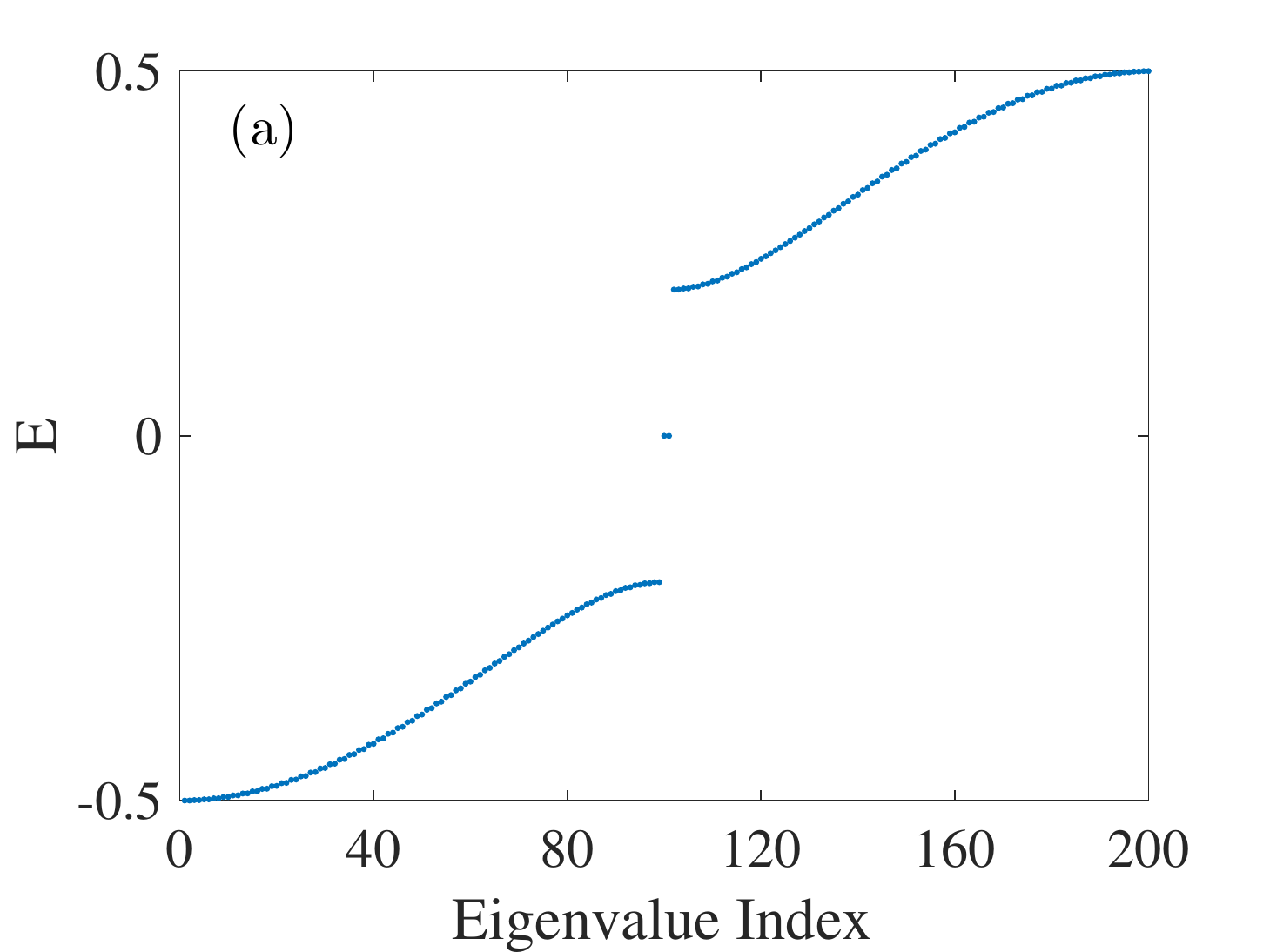}}
	\subfigure{\includegraphics[width=0.85\linewidth]{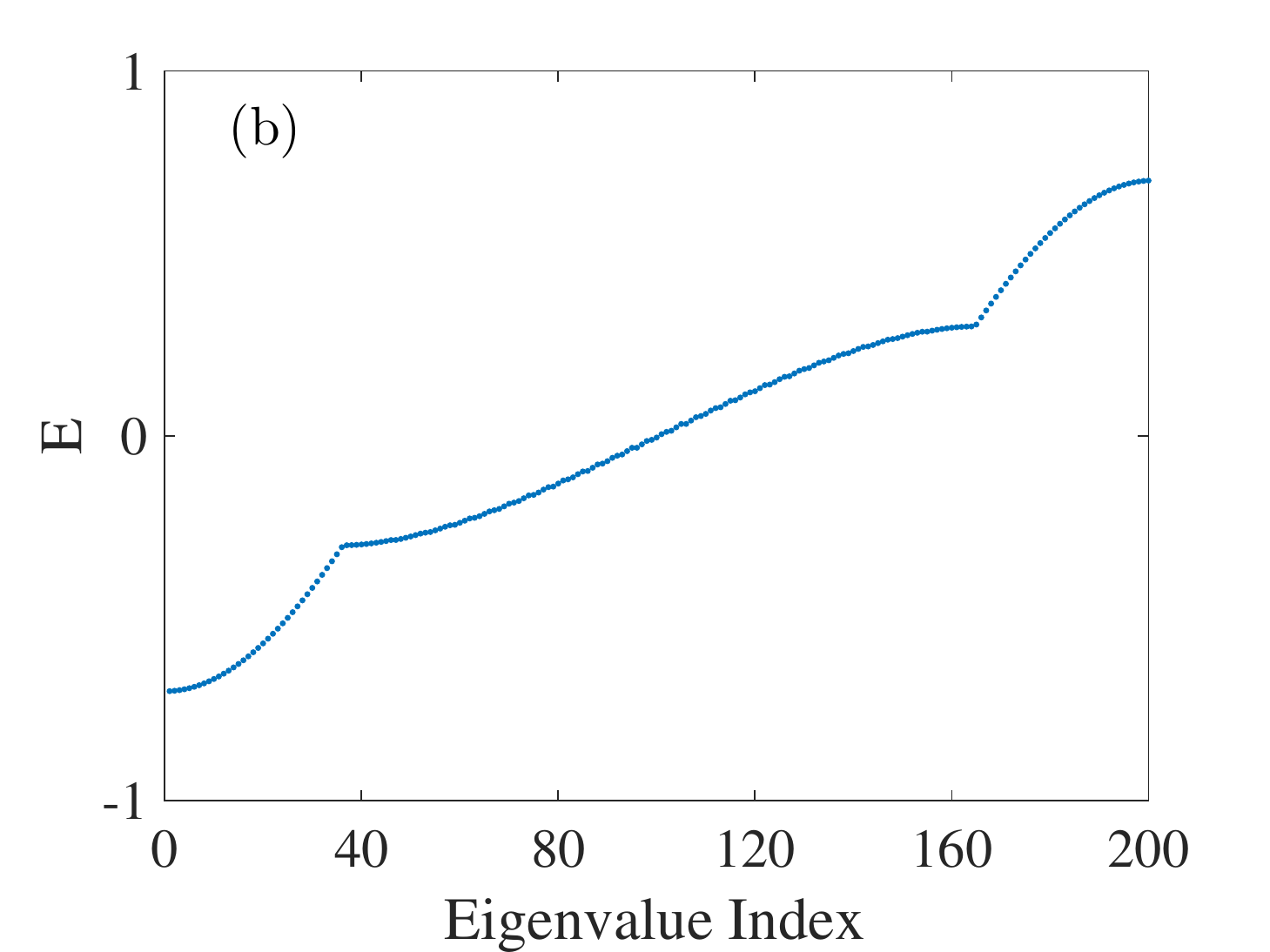}}
	\caption{The energy spectra ($2L$ eigenvalues) of the system under Majorana representation. (a) Standard Kitaev energy spectrum when the system satisfies Femi exchange antisymmetry ($G_{c}=0.5$, $G_{c}J_{-2}(\kappa_{1,n})=G_{c}J_{-2}(\kappa_{2,n})=0.2$). (b) Kitaev energy spectrum of the present system which does not satisfy Femi exchange antisymmetry ($G_{c}=0.5$, $G_{c}J_{-2}(\kappa_{1,n})=G_{c}J_{-2}(\kappa_{2,n})=0.2$). The size of the system is $L=100$.}\label{fig3}
\end{figure}

In this section, we will induce an analogous Kitaev Hamiltonian using the frequency modulations of cavity fields and resonators in the 1D multi-resonator optomechanical system. Consider the parameter regime of $\lambda_{n+1}=\lambda_{n}=\gamma_{n}$, $\phi \ne 0 $, and $T=0$, after performing the Jacobi$-$Anger transformation under the red-detuning regime again, the Hamiltonian of the system can be derived as
\begin{eqnarray}\label{e13}
H_{L,C_{1}}&&=\sum_{n} \left[-G_{n}a_{n}^{\dag}b_{n}+G_{n+1}a_{n+1}^{\dag}b_{n}\right.\cr\cr
&&\left.+\sum_{m_{1}=-\infty}^{\infty} -G_{n}J_{m_{1}}(\kappa_{1,n})a_{n}^{\dag}b_{n}^{\dag}e^{i(2\omega_{b}+m_{1}\nu) t}e^{im_{1}\phi}\right.\cr\cr
&&\left.+\sum_{m_{2}=-\infty}^{\infty}G_{n+1}J_{m_{2}}(\kappa_{2,n})a_{n+1}^{\dag}b_{n}^{\dag}e^{i(2\omega_{b}+m_{2}\nu) t}e^{im_{2}\phi}\right]\cr\cr
&&+\mathrm{H.c.}.
\end{eqnarray}
The resonant Stokes heating terms can be obtained by choosing $m_{1}=m_{2}=-2$. With the choice of $\phi=-0.25\pi$, then the Hamiltonian becomes
\begin{eqnarray}\label{e14}
H_{L,C_{2}}&=&\sum_{n} \left[-G_{n}a_{n}^{\dag}b_{n}+G_{n+1}a_{n+1}^{\dag}b_{n}\right.\cr\cr
&&\left.-iG_{n}J_{-2}(\kappa_{1,n})a_{n}^{\dag}b_{n}^{\dag}+iG_{n+1}J_{-2}(\kappa_{2,n})a_{n+1}^{\dag}b_{n}^{\dag}\right]\cr\cr
&&+\mathrm{H.c.}.
\end{eqnarray}  
For the mapping of Kitaev model Hamiltonian, we choose the effective optomechanical coupling strength to satisfy $-G_{n}=G_{n+1}=G_{c}$ (a fixed value). After that, the Hamiltonian becomes
\begin{eqnarray}\label{e15}
H_{L,C}&=&\sum_{n} \left[G_{c}a_{n}^{\dag}b_{n}+G_{c}a_{n+1}^{\dag}b_{n}\right.\cr\cr
&&\left.+iG_{c}J_{-2}(\kappa_{1,n})a_{n}^{\dag}b_{n}^{\dag}+iG_{c}J_{-2}(\kappa_{2,n})a_{n+1}^{\dag}b_{n}^{\dag}\right]\cr\cr
&&+\mathrm{H.c.}.
\end{eqnarray}  
One can clearly see that the above Hamiltonian has the identical form as the Kitaev model with zero chemical potential, in which the first two terms represent the NN hopping and the imaginary Stokes heating terms are analogous with the superconducting pairing terms of Kitaev model. We simulate the energy spectra of the standard fermionic Kitaev model and the present cavity optomechanical lattice chain system numerically, as shown in Figs.~\ref{fig3}(a) and \ref{fig3}(b). Compared to the standard fermionic Kitaev model, we find that the present system only possesses a continuous non-gapped energy spectrum corresponding to arbitrary values of $J_{-2}(\kappa_{1,n})$ and $J_{-2}(\kappa_{2,n})$, which is significantly different from the fermionic Kitaev energy spectrum exhibiting two degenerate zero energy modes in the gap. The reason is that the bosonic operators satisfy the commutation relation, which means that simply replacing the fermionic operators of standard Kitaev model with bosonic operators cannot obtain the analogous Majorana double chains sructure as the fermionic Kitaev model~\cite{mcdonald2018phase}. Therefore, the present bosonic Kitaev Hamiltonian with the same real-space form as fermionic Kitaev model exhibits the trivial topology and has a non-gapped energy spectrum.

\subsection{\label{sec.D}Effects of partial next-nearest-neighboring hopping regulated by frequency modulations}
\begin{figure}
	\centering
	\includegraphics[width=1\linewidth]{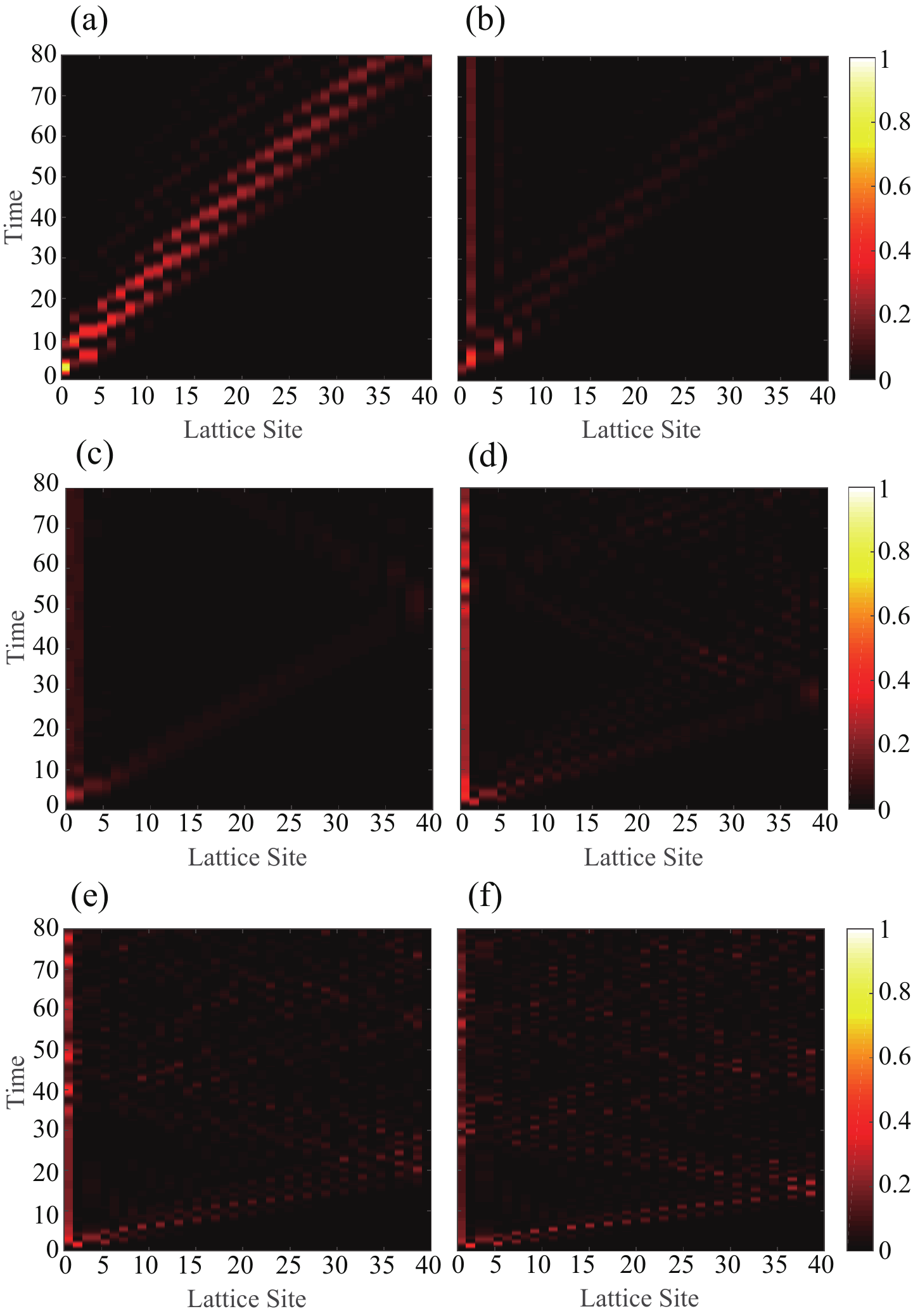}\\
	\caption{The effects of topological edge state and the partial NNN hopping on quantum walks. (a) Quantum walks in a nontopological multi-resonator optomechanical system ($-G_{n}>G_{n+1}$). (b) Quantum walks suppressed by topological edge states ($-G_{n}<G_{n+1}$). (c), (d), (e), and (f) The suppression of topological edge states are destroyed by the partial NNN hopping. In (c), (d), (e), and (f) the strengths of the partial NNN hopping are $TJ_{0}(\kappa_{1,n})=0.1,~0.45,~0.75,~1$ respectively. The size of the system is $L=40$.}\label{fig4}
\end{figure}
\begin{figure}
	\centering
	\includegraphics[width=1\linewidth]{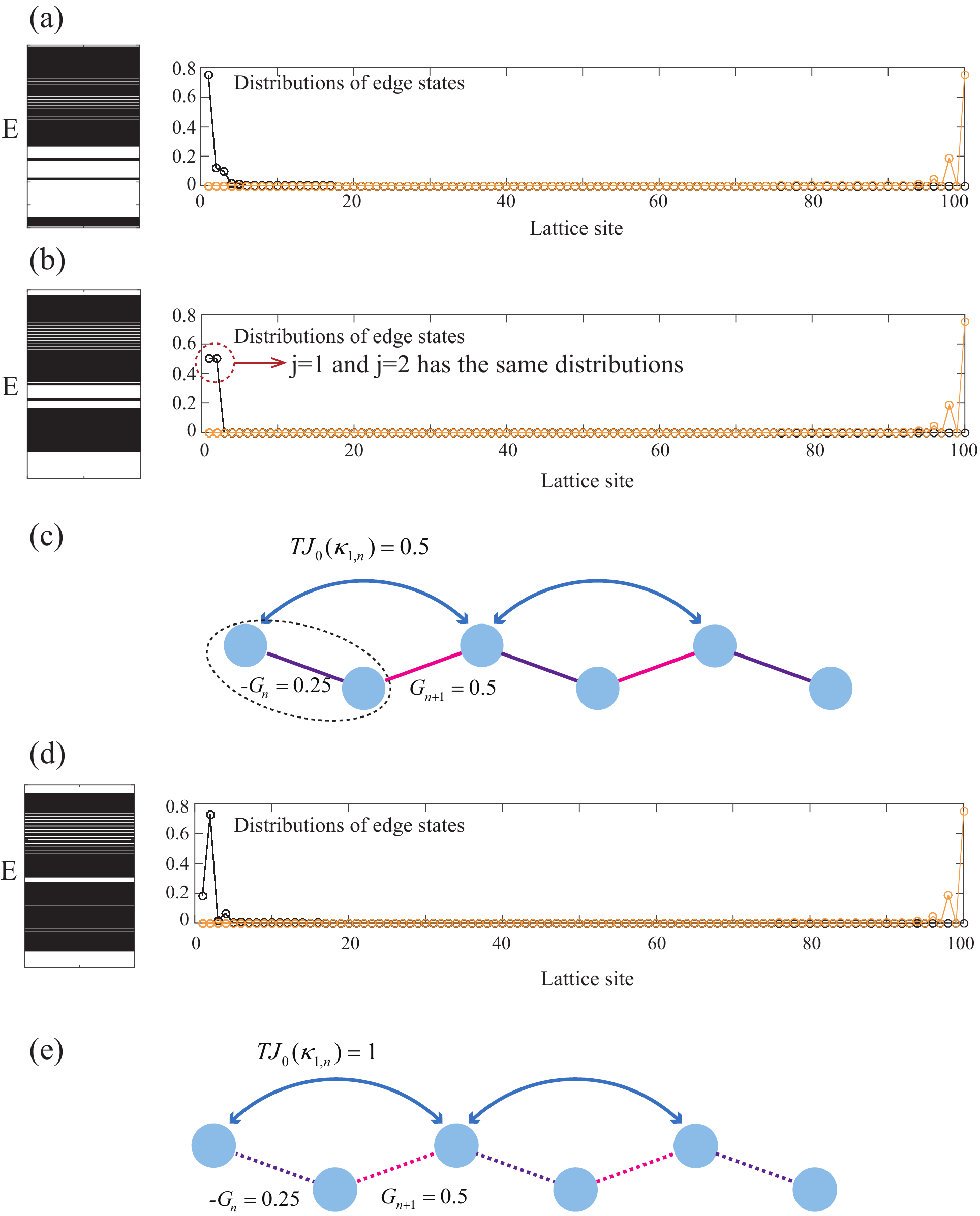}\\
	\caption{Energy spectra and distributions of edge states. (a) Energy spectrum and distributions of edge states corresponding to $T J_{0}(\kappa_{1,n})=0.25$. (b) Energy spectrum and distributions of edge states corresponding to $T J_{0}(\kappa_{1,n})=0.5$. (c) The schematic of the multi-resonators optomechanical system when the partial NNN coupling satisfies $T J_{0}(\kappa_{1,n})=G_{n+1}=0.5$. The black dashed circle represents the super-site. (d) Energy spectrum and distributions of edge states corresponding to $T J_{0}(\kappa_{1,n})=1$. (e) The schematic of the multi-resonator optomechanical system under the large partial NNN coupling limit $T J_{0}(\kappa_{1,n})=1$. The dashed lines represent the decoupling due to $T J_{0}(\kappa_{1,n})\gg G_{n+1}$. The quantum walk mainly pass though the odd sites (the path indicated by blue arrows). The other parameters are set as $-G_{n}=0.25$ and $G_{n+1}=0.5$. The size of the system is $L=100$.}\label{fig5}
\end{figure}
The interaction between two adjacent cavity fields is not taken into account in the above discussions. However, the realistic cavity optomechanical system may possess a strong direct coupling between cavity fields, which corresponds to the partial NNN hopping in the perspective of mapping the cavity field $a_{n}$ and resonator $b_{n}$ as two sites of the SSH model. In this way, the direct coupling between two cavity fields is equivalent to the NNN hopping only added onto the odd sites of the SSH chain. In the following we discuss the effects of partial NNN hopping on the system via choosing the parameters of frequency modulations to satisfy $\lambda_{n+1}=\lambda_{n}=\gamma_{n}$, $\phi = 0 $, and $T\ne 0$. After removing the Stokes heating terms as discussed in Sec.~\ref{sec.B}, the system can be dominated by the following tight-binding Hamiltonian  
\begin{eqnarray}\label{e16}
H_{L,D_{1}}&=&\sum_{n} \left[-G_{n}a_{n}^{\dag}b_{n}+G_{n+1}a_{n+1}^{\dag}b_{n}\right.\cr\cr
&&\left.+\sum_{m=-\infty}^{\infty} T J_{m}(\kappa_{1,n})a_{n+1}^{\dag}a_{n}e^{i m\nu t}\right]\cr\cr
&&+\mathrm{H.c.},
\end{eqnarray} 
where $\kappa_{1,n}=\lambda_{n+1}-\lambda_{n}$. When the first kind of Bessel function takes the zeroth order ($m=0$), the system possesses the resonant partial NNN hopping. The Hamiltonian is simplified as
\begin{eqnarray}\label{e17}
H_{D}&=&\sum_{n} \left[-G_{n}a_{n}^{\dag}b_{n}+G_{n+1}a_{n+1}^{\dag}b_{n}+ T J_{0}(\kappa_{1,n})a_{n+1}^{\dag}a_{n}\right]\cr\cr
&&+\mathrm{H.c.},
\end{eqnarray} 
where $T J_{0}(\kappa_{1,n})$ represents the effective partial NNN hopping strength regulated by Bessel function. The multi-resonator optomechanical system can be regarded as a standard SSH chain when the parameters satisfy $J_{0}(\kappa_{1,n})=0$ and $-G_{n}<G_{n+1}$. We stress that the condition of $\lambda_{n+1}=\lambda_{n}$ means $J_{0}(\kappa_{1,n})=1$ corresponding to a non-regulated partial NNN hopping. However, the realistic selections of parameters may possess the relative fluctuations, which leads $J_{0}(\kappa_{1,n})\ne 0$ corresponding to a regulated partial NNN hopping. 

The mapping of topological SSH model based on a 1D multi-resonator optomechanical system provides a new kind of optical platform to investigate various effects of topology on other physics, such as the block of topological edge states on quantum walks. As shown in Fig.~\ref{fig4}, we investigate the influences of topological edge states and partial NNN hopping on quantum walks. The numerical results reveal that the appearance of topological edge states will have a suppression effect on quantum walks in the present system, as shown in Figs.~\ref{fig4}(a) and \ref{fig4}(b). However, when the partial NNN hopping is introduced into the system, we find that the existence of partial NNN hopping will destroy the suppression of the topological edge states on quantum walks with the raising of the partial NNN hopping strength, as shown in Figs.~\ref{fig4}(c)-\ref{fig4}(f). The reason is that the large partial NNN interactions will accelerate the process of quantum walks in the bulk.

More specifically, when the parameter satisfies $-G_{n}<G_{n+1}$ (topologically nontrivial), the topological left edge state has a maximal suppression effect on quantum walks corresponding to $T J_{0}(\kappa_{1,n})=0$. The two degenerate zero energy modes are separated accompanying with the decrease of the distribution of left edge state when a mild partial NNN hopping ($T J_{0}(\kappa_{1,n})<-G_{n}$) is added on the system, which leads a weak suppression effect of left edge state on quantum walks, as shown in Fig.~\ref{fig5}(a). With the strength of the partial NNN hopping continuously increasing, the distribution of left edge state decreases continuously while the distribution of the second site enlarges gradually, which further weakens the suppression effect of left edge state on quantum walks. A phase transition occurs when the partial NNN hopping strength reaches the same value as the strength of the NN hopping $G_{n+1}$ ($T J_{0}(\kappa_{1,n})=G_{n+1}$), in which the first cavity field and the second resonator have the same distributions, as shown in Fig.~\ref{fig5}(b). The reason of this phenomenon is that the first cavity field and the second resonator will be equivalent to a super-site due to the same effective hopping strength $T J_{0}(\kappa_{1,n})=G_{n+1}$, as shown in Fig.~\ref{fig5}(c). It means that quantum walks can pass though the lattice in terms of two paths (the path indicated by blue arrows and the path indicated by pink lines, respectively) with the same probability at the same time. After that, the distribution of the second site raises continuously with the sequential increase of the partial NNN hopping strength, and a large enough partial NNN hopping strength ($T J_{0}(\kappa_{1,n})\gg G_{n+1}$) will separate the second sites from the lattice chain, which generates a new localized state, as shown in Fig.~\ref{fig5}(d). The large enough partial NNN hopping limit means that the quantum walks will pass though the lattice only in terms of the odd sites, which leads the acceleration of quantum walks and destroys the suppression effect of left edge state on quantum walks due to the disappear of left edge state, as shown in Fig.~\ref{fig5}(e). These results enlighten that we can realize the controllable suppression and acceleration of quantum walks based on the present cavity optomechanical system with frequency modulations. 

\section{\label{sec.4}Conclusions}
In conclusion, we have proposed a scheme to induce the topological SSH phase via distinguishing parameters regimes based on a 1D multi-resonator cavity optomechanical system with frequency modulations. We find that, after eliminating the Stokes heating terms, the optomechanical chain system will exhibit a topological SSH phase by varying the effective NN hopping assisted by the frequency modulations alternately. The property of  Bessel function guarantees the realization of SSH model corresponding to arbitrary effective optomechanical coupling strength in the certain range. Another parameter regime is proposed to remove the resonant Stokes heating terms, in which the optomechanical chain system exhibits a SSH phase via changing the effective optomechanical coupling alternately. We also realize a bosonic Kitaev model with trivial topology by keeping the Stokes heating terms. Our scheme provides a controllable method to investigate the effect of the partial NNN hopping on the SSH model based on a 1D multi-resonator cavity optomechanical system with frequency modulations.

\begin{acknowledgments}
This work was supported by the National Natural Science Foundation of China under Grant Nos.
61822114, 61465013, and 11465020, and the Project of Jilin Science and Technology Development for Leading Talent of Science and Technology Innovation in Middle and Young and Team Project under Grant No. 20160519022JH.
\end{acknowledgments}


\section{Appendix: LINEARIZING THE SYSTEM HAMILTONIAN CORRESPONDING TO AN EVEN NUMBER OF LATTICE SITES}

Here, we consider the frequency-modulated optomechanical array composed of $N$ cavity fields and $N$ resonators (the lattice size is $2N$ corresponding to an even number), in which the coupling between the resonator $b_{n}$ and adjecant cavity field $a_{n}$ ($a_{n+1}$) is $g_{n}$. The system can be dominated by 
\begin{eqnarray}\label{e18}
H_{\mathrm{even}}&=&\sum_{n=1}^{N}\big\{\left[\omega_{a,n}+\lambda_{n} \nu\cos(\nu t+\phi)\right]a_{n}^{\dag}a_{n}\cr\cr
&&+(\Omega_{n}a_{n}^{\dag}e^{-i\omega_{d}t}+\Omega_{n}^{\ast}a_{n}e^{i\omega_{d}t})\cr\cr
&&+\left[\omega_{b,n}+\gamma_{n} \nu\cos(\nu t+\phi)\right]b_{n}^{\dag}b_{n}\cr\cr
&&-g_{n} a_{n}^{\dag}a_{n}(b_{n}^{\dag}+b_{n})\big\}\cr\cr
&&+\sum_{n=1}^{N-1} g_{n}a_{n+1}^{\dag}a_{n+1}(b_{n}^{\dag}+b_{n}).
\end{eqnarray}

Under the condition of strong laser driving, we perform the standard linearization approach to linearize the Hamiltonian. After dropping the notation ``$\delta$'' for all the fluctuation operators $\delta a_{n}$ ($\delta b_{n}$), the Hamiltonian can be deformed as
\begin{eqnarray}\label{e19}
H_{L}&=&\sum_{n=1}^{N}\big\{[\Delta_{a,n}^{'}+\lambda_{n} \nu\cos(\nu t+\phi)]a_{n}^{\dag}a_{n}\cr\cr
&&+\left[\omega_{b,n}+\gamma_{n} \nu\cos(\nu t+\phi)\right]b_{n}^{\dag}b_{n}\cr\cr
&&-g_{n} (\alpha_{n}^{\ast}a_{n}+\alpha_{n}a_{n}^{\dag})(b_{n}^{\dag}+b_{n})\big\}\cr\cr
&&+\sum_{n=1}^{N-1} g_{n}(\alpha_{n+1}^{\ast}a_{n+1}+\alpha_{n+1}a_{n+1}^{\dag})(b_{n}^{\dag}+b_{n}),\cr
&&
\end{eqnarray}
where $\Delta_{a,n}^{'}$ contains $\Delta_{a,1}^{'}=\Delta_{a,1}+g_{1}(\beta_{1}^{\ast}+\beta_{1})$ and $\Delta_{a,n=2...N}^{'}=\Delta_{a,n}-g_{n-1}(\beta_{n-1}^{\ast}+\beta_{n-1})+g_{n}(\beta_{n}^{\ast}+\beta_{n})$. Performing a rotating transformation on the linearization Hamiltonian in Eq.~(\ref{e19}) with
\begin{eqnarray}\label{e20}
V&=&\mathrm{exp}\bigg\{\sum_{n=1}^{N}-i\Delta_{a,n}^{'}ta_{n}^{\dag}a_{n}-i\lambda_{n}\sin(\nu t+\phi) a_{n}^{\dag}a_{n}\cr\cr
&&-i\omega_{b,n} t b_{n}^{\dag}b_{n}-i\gamma_{n} \sin(\nu t+\phi) b_{n}^{\dag}b_{n}\bigg\},
\end{eqnarray}
then the Hamiltonian becomes
\begin{widetext}
\begin{eqnarray}\label{e21}
H_{L}^{'}&=&\sum_{n=1}^{N} \left\{-G_{n}a_{n}^{\dag}b_{n}e^{i\left[(\Delta_{a,n}^{'}-\omega_{b,n})t+(\lambda_{n}-\gamma_{n})\sin(\nu t+\phi) \right]}-G_{n}a_{n}^{\dag}b_{n}^{\dag}e^{i\left[(\Delta_{a,n}^{'}+\omega_{b,n})t+(\lambda_{n}+\gamma_{n})\sin(\nu t+\phi) \right]}\right\}\cr\cr
&&+\sum_{n=1}^{N-1} \left\{G_{n+1}a_{n+1}^{\dag}b_{n}e^{i\left[(\Delta_{a,n+1}^{'}-\omega_{b,n})t+(\lambda_{n+1}-\gamma_{n})\sin(\nu t+\phi) \right]}+G_{n+1}a_{n+1}^{\dag}b_{n}^{\dag}e^{i\left[(\Delta_{a,n+1}^{'}+\omega_{b,n})t+(\lambda_{n+1}+\gamma_{n})\sin(\nu t+\phi) \right]}\right\}\cr\cr
&&+\mathrm{H.c.},
\end{eqnarray}
\end{widetext}
where $G_{n}=g_{n} \alpha_{n}$ ($G_{n+1}=g_{n} \alpha_{n+1}$) is the effective optomechanical coupling parameter. Obviously, the above Hamiltonian has the same form as the Hamiltonian in Eq.~(\ref{e04}). Thus, our scheme is also valid for the case of an even number of lattice sites. To avoid the tedious expression of two summation symbols in Eq.~(\ref{e21}), we focus on the case of an odd number of lattice sites in the main text.



\end{document}